\documentclass[twocolumn,showpacs,preprintnumbers,amsmath,amssymb]{revtex4}
\usepackage{epsfig}

\newcommand{\etal}[1]{{\it et al.}}

\newcommand{\HHCO}{H$_2$CO}
\newcommand{\NDDD}{ND$_3$}
\newcommand{\NHHH}{NH$_3$}
\newcommand{\vt}{v_t}
\newcommand{\vl}{v_l}
\newcommand{\vtmax}{{v_{t {\rm max}}}}
\newcommand{\vlmax}{{v_{l {\rm max}}}}

\begin{document}

\title{A continuous source of translationally cold dipolar molecules}

\author{S.A. Rangwala, T. Junglen, T. Rieger, P.W.H. Pinkse, and G. Rempe}
\affiliation{Max-Planck-Institut f\"ur Quantenoptik,  Hans-Kopfermann-Str. 1,
D-85748 Garching, Germany}
\date{\today, PREPRINT}

\begin{abstract}
The Stark interaction of polar molecules with an inhomogeneous
electric field is exploited to select slow molecules from a
room-temperature reservoir and guide them into an ultrahigh vacuum
chamber. A linear electrostatic quadrupole with a curved section
selects molecules with small transverse and longitudinal
velocities. The source is tested with formaldehyde (\HHCO) and
deuterated ammonia (\NDDD). With \HHCO\ a continuous flux is
measured of $\approx 10^9$/s and a longitudinal temperature of a
few K. The data are compared with the result of a Monte Carlo
simulation.


\end{abstract}
\maketitle

\vspace{1cm}


The past years have seen an explosion of activity in the field of
cold atomic gases~\cite{Nobelrefs}. It is interesting and
desirable to extend these investigations to molecules, which have
a complex internal structure and can as a consequence possess a
permanent electric dipole moment. Trapping cold polar molecules
will lead to new physics due to the long range and anisotropy of
the dipole-dipole interaction. Slow molecules for precision
measurements or interferometry are further motivations behind the
ongoing efforts. However, the complexity and density of energy
levels in the rotational and vibrational manifolds largely
precludes the effective use of laser cooling
techniques~\cite{BahnsGouldStwalley}.
Therefore, a number of different approaches has been considered
for cooling and trapping molecules. Buffer-gas cooling in a
cryogenic environment is one possibility, but requires a rather
complex setup~\cite{WeinsteinNature98}. Another method is
photoassociation, but this is limited to simple molecules with
laser-cooled precursor atoms~\cite{Photoassociationrefs}. A novel
technique uses deceleration by the Stark effect, where packages of
polar molecules are decelerated with time-varying electric
fields~\cite{BethlemPRL99,MaddiPRA99,BethlemNature00}. Other,
mostly mechanical methods have also been proposed but remain to be
demonstrated~\cite{GuptaJCP99,FriedrichPRA61}.

It is, however, not necessary to {\em produce} slow molecules, as
they are present in any thermal gas, even at room temperature.
Slow molecules only need to be filtered out. For this reason,
already in the 1950s it was attempted to select the slowest atoms
from a hot beam using gravity~\cite{Zacharias}. These attempts
failed, mostly because the slow particles were kicked away by the
fast ones. Much later, it was demonstrated that slow lithium atoms
can be efficiently guided out of a hot beam with strong permanent
magnets, providing a robust and cheap source of slow atoms
~\cite{GhaffariPRA99}, e.g. for Bose-Einstein condensation
experiments. In the same spirit, an efficient and simple filtering
technique could play an important role towards the production of a
cold molecular gas.

In this Letter we describe an experiment in which the Stark
interaction of polar molecules with an inhomogeneous,
electrostatic field is exploited to efficiently select and guide
slow molecules out of a room-temperature reservoir into ultrahigh
vacuum. Whether a dipolar molecule is weak-field seeking and
trapped by an electric field minimum, or strong-field seeking and
expelled, depends on whether the average orientation of the
rotating molecular dipole is antiparallel or parallel to the local
electric field, respectively. Weak-field seeking molecules are
trapped if their kinetic energy is less than the Stark-potential
barrier for them. Thus a static quadrupolar potential forms a
two-dimensional trap for molecules with small transverse velocity
components with respect to the quadrupole axis. In a linear guide,
transverse and longitudinal components of the velocity are
completely decoupled. The longitudinal velocity of the molecules
is limited by guiding them around a bend in the quadrupole,
downstream from the reservoir. The centripetal force due to the
electric field gradient guides only the slowest molecules around
the bend. The fast molecules escape the guide and are pumped away.

Important parameters for characterizing such a continuous source of slow
molecules are the velocity distribution and the flux of the molecules. For a
given molecular state, we have a maximum transverse ($\vtmax$) and longitudinal
($\vlmax$) velocity that will be guided. Under the assumptions that the Stark
shift of the molecule is linear with electric field and that the longitudinal
($\vl$) and transverse ($\vt$) velocities are much smaller than the mean
thermal velocity inside the reservoir, the flux $\Phi$ in the guide is
\begin{equation}
    \Phi\propto\int_{\vt=0}^\vtmax
    2\pi \vt\, {\rm d}\vt \int_{\vl=0}^\vlmax \vl\, {\rm d}\vl\propto v_{t {\rm
    max}}^2 v_{l {\rm max}}^2\propto E^2,
    \label{Eq:Phi}
\end{equation}
where $E$ is the depth-determining electric field strength.
Eq.~(\ref{Eq:Phi}) is valid for every molecule with a linear Stark
state, and therefore also for an ensemble of molecules with
different, but linear, Stark shifts. The derivation of
Eq.~(\ref{Eq:Phi}) utilizes a unique property of the guide,
namely, that the longitudinal velocity distribution has a linear
dependence on velocity for small velocities: $\Phi(\vl)\propto\vl
\exp[-\vl^2/\alpha^2]$, where $\alpha$ is the most probable
velocity for molecules in the source, in contrast to molecular
beams, which have a $\vl^3 \exp[-\vl^2/\alpha^2]$ dependence. The
reason is that the guide selects on kinetic energy, not on angle
as in a typical molecular beam, which is collimated by apertures.
As a consequence, molecules that are slow in axial and radial
direction will still be guided, whereas they would be lost in a
molecular beam.


\begin{figure}[htb]\begin{center}
\epsfig{file=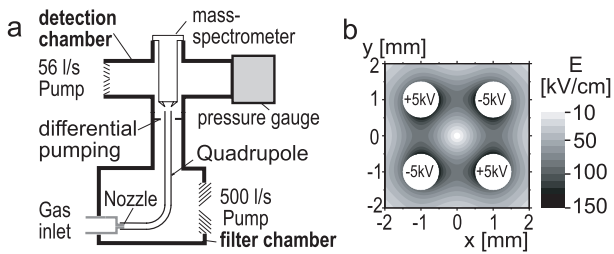,width=0.45\textwidth}%
\caption{(a) Schematic of the experimental setup, and (b) a
contour plot of the quadrupolar electric field in a plane
perpendicular to the electrode rods. } \label{Fig.Setup}
\end{center}\end{figure}

Most of our experiments are performed with formaldehyde (\HHCO), which has a
relatively large dipole moment of 2.34\,Debye and is one of the best studied
4-atomic asymmetric-top molecules. At room temperature (300\,K) almost all the
\HHCO\ molecules will be in the vibrational ground state. A large number of
rotational states consistent with the Boltzmann distribution at 300\,K are
populated. The experiment itself (Fig.~\ref{Fig.Setup}(a)) consists of the bent
quadrupolar guide starting at a $0.5\,$mm diameter ceramic nozzle of $11\,$mm
length and ending in a UHV detection chamber, passing through a vacuum chamber
where the filtering is performed. The guide has a length of 18\,cm and is made
of 1\,mm diameter stainless steel rods, with a 1\,mm gap between neighboring
rods. The rods are built around the ceramic nozzle. Typical operation pressures
in the nozzle, which injects the molecules directly into the quadrupole, are
below 0.05\,mbar in order to maintain molecular flow conditions. Most of the
molecules are not guided and escape into the vacuum chamber, where a typical
operational pressure of a few times $10^{-7}\,$ mbar is maintained by a
500\,l/s turbo-molecular pump. The $90^\circ$ bent section of the guide is
about 60\,mm downstream from the nozzle and its radius of curvature is
13.5\,mm. After the bend, the guide passes through a 5\,cm long and 4\,mm
narrow tube for differential pumping before entering the detection chamber,
where a 56\,l/s turbo pump maintains a pressure of $\approx2\times10^{-9}\,$
mbar. The guide abruptly ends 25\,mm before the ionization volume of a
quadrupole mass spectrometer (QMS) [Pfeiffer vacuum, Prisma QMA 200]. Apart
from measuring the direct flux, the QMS monitors the residual gas in the
detection chamber. Standard pressure gauges are used to monitor the pressure in
the UHV chamber.

\begin{figure}[htb]\begin{center}
\epsfig{file=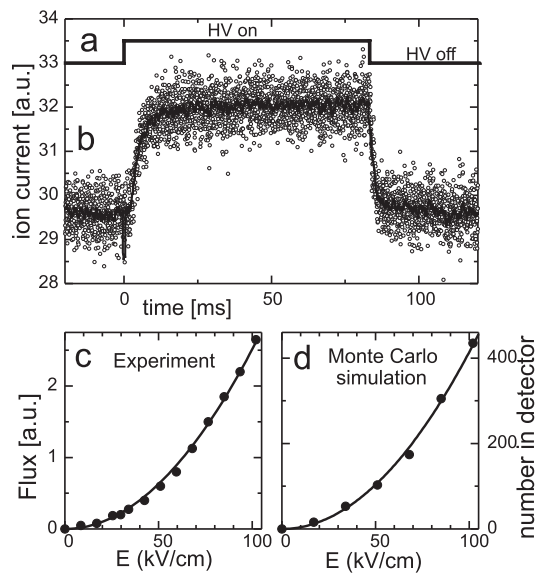,width=0.4\textwidth}%
\caption{(a) The high voltage applied to the guide electrodes and (b) the
molecular flux as a function of time. Note the delay and the relatively slow
rise of the signal, corresponding to a gradual build-up of molecular flux,
compared to the sudden fall due to the loss of molecules when the field is
switched off. The line is a 20-point running average. (c) The flux as a
function of applied electrode voltage. The symbols denote the measured height
of the step function in (b). (d) The number of trajectories intersecting with
the detector in a simulation where $10^6$ particles were injected into the
guide, corresponding to a 1\% fraction of the full input distribution. The
lines in (c) and (d) are quadratic fits. } \label{Fig.Signal}
\end{center}\end{figure}

To interpret experimental results, a Monte Carlo simulation was
performed. The electric field in a $90^{\circ}$-bent quadrupole
with short straight sections on either side was calculated
numerically in 3 dimensions. The input distribution of the
molecules injected into the guide in the simulation is random and
consistent with the thermal speed distribution. Its angle
distribution is that of a $3\,$mm deep field-free
nozzle~\cite{NozzleAngleDistriRef}. This is a good approximation
to the experimental situation, where a $11\,$mm long nozzle in the
presence of an electric field results in the transverse
acceleration of molecules, giving a wall-to-wall collision-free
path length of approximately $3\,$mm for the guidable molecules.
For each molecular species, the Stark shifts of the relevant
states were calculated by direct numerical diagonalization of the
Stark Hamiltonian~\cite{HainJCP99}. In the relevant electric field
range (0-100\,kV/cm), the majority of thermally populated
rotational states of \HHCO\ and \NDDD\ show linear Stark shifts.
These were gathered in a small number (20) of groups with similar
Stark shifts. The trajectories under the influence of the
electrostatic forces were calculated by a Runge-Kutta method. The
simulation confirmed the expected quadratic dependence of the flux
on the electric field, see Fig.\ref{Fig.Signal}(d). The simulation
also yields the overall transmission of the guide, the velocity
distribution inside the guide, the output divergence and velocity
distribution of the beam. Finally, the simulation shows that the
guide enriches the gas with states having a large Stark shift.


For detecting \HHCO\ molecules the quadrupole mass filter is set
at mass 29, the strongest peak in the \HHCO\ mass spectrum. The
influence of other gases at this mass is negligible. The
channeltron-amplified QMS ion current can be tapped directly for
transient measurements. Switching on and off the quadrupolar field
resulted in a modulated QMS signal, see Fig.~\ref{Fig.Signal}(a)
and (b). It was checked that the electric disturbance from
switching the high voltage did not significantly influence the
signal apart from a short spike of sub micro-second duration. At
high background pressures of polar molecules in the filter chamber
and without injection of polar molecules into the guide, the ion
signal {\em decreases} slightly when the HV is turned on. This
effect is attributed to a decrease of conductance of the
differential pumping section for polar molecules in the presence
of electric fields. The effect is absent for non-polar molecules.
As a further test that the signal represents the direct flux of
the guide, a mechanical shutter was installed that can prevent the
direct flux from reaching the QMS. It is designed so that it only
prevents the guided molecules from entering the QMS. With the
shutter blocking the direct flux into the QMS, the modulation of
the QMS signal was a factor of $\approx5$ smaller. With the above
tests we conclude that the observed changes in the QMS signal are
due to real changes in the density of the guided gas.

\begin{figure}[htb]\begin{center}
\epsfig{file=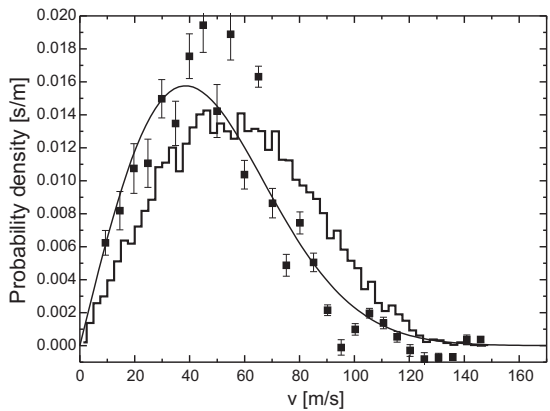,width=0.4\textwidth}%
\caption{The longitudinal velocity distribution (data points with statistical
error bars) derived from data obtained with an electrode voltage of $\pm5\,$kV.
The smooth curve is a (normalized) fit to the data of the functional form
$(2\vl/\alpha^2) \exp[-\vl^2/\alpha^2]$, with $\alpha=54\,$m/s, the stepped
curve is the Monte-Carlo results for 5 kV. The negative values at high
velocities are due to statistical noise in the data at short times. }
\label{Fig.Veldistri}
\end{center}\end{figure}

As a first measurement, the guiding signal was recorded as a
function of the applied voltage. The result, shown in
Fig.~\ref{Fig.Signal}(c), shows the expected quadratic dependence.
To further characterize the source, the rising slope of the
signal, coming from the time-of-flight, was analyzed in more
detail. By using fast switches, the turn-on time of the high
voltage is less than $1\,\mu$s, much less than all other relevant
time scales in the experiment. With the high voltage off, the
density of slow molecules that could be guided decays rapidly with
distance from the nozzle. Therefore, it can be assumed that the
molecules arriving at the detector must have entered the guide
after the high voltage has been switched on. After a delay of a
few ms, which depends on the applied voltage, the fastest guided
molecules arrive at the QMS and the signal starts rising until it
levels off after $\approx10\,$ms. From the delay and the rising
slope, the longitudinal velocity distribution can be derived by
differentiation. The analysis incorporates the following
additional inputs : 1) the Monte Carlo simulation was used to
estimate the velocity-dependent probability of hitting the
detector after emerging from the guide. This probability was used
to perform a velocity-dependent correction, which accounts for the
acceleration of the molecules when leaving the guide and the fact
that longitudinally slow molecules are more likely to miss the
detector, because they spread out over a larger solid angle; 2)
making the reasonable assumption that the ionizing probability of
the QMS depends on the molecular velocity and does not saturate at
the lowest measured velocities, a velocity-dependent ionization
probability was calculated. The resulting velocity distribution
obtained for an electrode voltage of $\pm5\,$kV is shown in
Fig.~\ref{Fig.Veldistri}. Subtracting the signal with the shutter
closed had only a small effect on the velocity distribution, and
was omitted.

Note that for a single molecular state the velocity distribution
should show a relatively sharp velocity cut-off. But given a
mixture of states with different Stark shifts, the cut-off is
smeared out. Surprisingly, this leads to a velocity distribution
which can reasonably well be described by a 1-dimensional thermal
distribution. For example, the smooth solid curve in
Fig.~\ref{Fig.Veldistri} represents a fit to the data with a
temperature of $5.4\,$K. The experimentally determined velocity
distribution is slightly narrower than the one from the
Monte-Carlo simulation, and the mean velocity is shifted to a
smaller value. This deviation is probably caused by experimental
imperfections such as surface roughness of the electrodes,
deviations from the design geometry, etc. The transverse velocity
distribution of the guided molecules could not be measured
directly, however, simulations indicate a distribution
characterized by a temperature of $0.5$\,K inside the guide for
$\pm5\,$kV electrode potential.

\begin{figure}[htb]\begin{center}
\epsfig{file=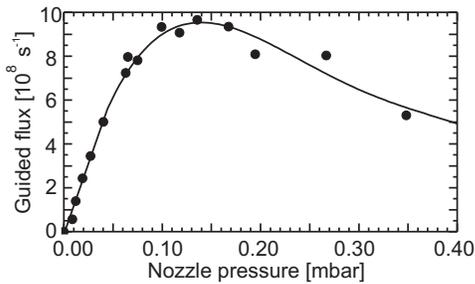,width=0.35\textwidth}%
\caption{The absolute measured flux through our guide as a function of nozzle
pressure for an electrode voltage of $\pm5\,$kV. The symbols are data points,
the curve a guide to the eye. For low pressures the gas in the nozzle is in the
molecular flow regime and the flux is linear in the nozzle pressure. Above
$\approx0.02\,$mbar the mean free path for thermal molecules becomes less than
the nozzle length and collisions start playing a role. As a consequence, the
increase of the flux slows down and eventually the absolute guided flux is
reduced.} \label{Fig.Flux}
\end{center}\end{figure}

The absolute flux was measured by calibrating the ion current from
the QMS with the background signal caused by non-guided hot \HHCO\
molecules. The partial pressure of the hot \HHCO\ was measured
with the QMS and related to an absolute pressure via a Penning
pressure gauge [Balzers, compact full range gauge PKR260], taking
into account the relative ionization probabilities from
Ref.~\cite{NISTHiden}. An ionization gauge [Varian UHV-24]
confirmed the Penning gauge readings to within 50\%. The
ionization volume is modelled by a $(3\,$mm$)^3$ box, as indicated
by the QMS supplier. Unfortunately, the exact dimensions of the
ionization volume are hard to verify for our conditions.
Therefore, we estimate a systematic error in the absolute flux of
the order of a factor of 2. The flux that was calibrated in this
way is plotted in Fig.~\ref{Fig.Flux} as a function of the
absolute nozzle pressure. The measured \HHCO\ content in the
nozzle is approximately 12\,\%. In Fig.~\ref{Fig.Flux}, the nozzle
pressure was deliberately increased to a point where the molecular
flow approximation is no longer valid. As can be seen in the
figure, for low nozzle pressures the flux increases linearly, but
starts deviating from the linear dependence above 0.05\,mbar,
reaches a maximum of nearly $10^9\,$s$^{-1}$, with peak density
inside the guide of $\approx10^8\,$cm$^{-3}$, around 0.15\,mbar
before actually reducing with pressure. This behavior is caused by
collisions. When increasing the pressure in the nozzle, the beam
characteristics change from an effusive, thermal velocity
distribution to a distribution which is peaked to a higher
velocity. Moreover, slow molecules are kicked away by the fast
ones in the guide and by background molecules. Therefore, it is
important to keep the pressure in the nozzle sufficiently low.

>From the Monte-Carlo simulation, the measured \HHCO\ concentration and the
theoretical flow impedance of the nozzle, the expected flux can be determined
at a nozzle pressure of $0.025\,$mbar and an electrode voltage of $\pm5\,$kV.
The result is $3.5\times10^{9}\,$s$^{-1}$, a factor of 10 larger than observed
in the experiment. The discrepancy might be due to background collisions and,
more likely, a non-perfect nozzle and a non-perfect alignment of the detector
and the electrodes.

To demonstrate the general character of the source, also a slow beam of
deuterated ammonia (\NDDD) was produced, with similar results to those obtained
for \HHCO. The deuterated species was used because of a null QMS background
signal from contaminants at the mass peaks of interest and because its Stark
shift is larger than that of \NHHH.

In conclusion, we have experimentally demonstrated a continuous, high-flux
source of translationally cold molecules at temperatures of a few kelvin. The
molecules are conveniently delivered ``on spot'', e.g. into an ultra-high
vacuum, where they can be stored in an electrostatic trap or an electrostatic
ring~\cite{CrompvoetsNature01}. This simple and versatile method is generally
applicable to all molecules with a reasonably large dipole moment. By reducing
the radius of curvature of the bend, slower output beams can easily be
achieved. In principle, the source can also be used as a novel kind of gas
chromatograph that selects polar molecules from a non-polar buffer gas, or for
enrichment of a gas with strong low-field-seeking states.

We thank W. Demtr\"oder and J. H\"ager for useful discussions, and
J. Bulthuis for checking the Stark shift calculations. Financial
support by the DFG is kindly acknowledged.


\end{document}